\documentclass[aps,prb,twocolumn,showpacs,preprintnumbers,amsmath,amssymb,superscriptaddress]{revtex4}

\usepackage{graphicx}
\usepackage{dcolumn}
\usepackage{bm}
\usepackage{tabularx}
\usepackage{multirow}
\newcommand{\nc}{\newcommand}
\nc{\be}{\begin{equation}}
\nc{\ee}{\end{equation}}
\nc{\bea}{\begin{eqnarray}}
\nc{\eea}{\end{eqnarray}}
\nc{\bean}{\begin{eqnarray*}}
\nc{\eean}{\end{eqnarray*}}
\nc{\mb}{\mbox}
\nc{\rnc}{\renewcommand}
\nc{\vk}{\mb{\bf k}}
\nc{\vp}{\mb{\bf p}}
\nc{\vn}{\mb{\bf n}}
\nc{\vq}{\mb{\bf q}}
\nc{\rr}{\mb{\bf r}}
\nc{\vz}{\hat {\mb{\bf z}}}
\nc{\vj}{\mb{\boldmath$j$}}
\nc{\vg}{\mb{\boldmath$g$}}
\nc{\x}{\mb{\boldmath$x$}}
\nc{\A}{\mb{\boldmath$A$}}
\nc{\va}{\mb{\boldmath$a$}}
\nc{\vs}{\mb{\boldmath$\sigma$}}
\nc{\vpi}{\mb{\boldmath$\pi$}}
\nc{\nab}{\nabla}
\nc{\X}{\sf x}
\nc{\asp}{\hspace{10mm}}

\begin{document}

\title{Theory of the SrTiO$_3$ Surface State Two-Dimensional Electron Gas}
\author{Guru Khalsa}
\author{ A.H. MacDonald}
\affiliation{Department of Physics, University of Texas at Austin, Austin TX 78712-1081, USA}
\date{\today}
\begin{abstract}
We present a theory of the quasi two-dimensional electron gas (2DEG) systems 
which appear near the surface of SrTiO$_3$ when a large external electric field 
attracts carriers to the surface.  We find that non-linear and non-local 
screening by the strongly polarizable SrTiO$_3$ lattice plays an essential role in 
determining 2DEG properties.  The electronic structure always includes 
weakly bound bulk-like bands that extend over many SrTiO$_3$ layers.
At 2D carrier-densities exceeding $\sim 10^{14} {\rm cm}^{-2}$ tightly bound
bands emerge that are confined within a few layers of the surface.
\end{abstract}
\maketitle

\section{Introduction} 
Two-dimensional electron gases can be formed in SrTiO$_3$ crystals\cite{Hwang_LAOSTO,Thiel,OxideInterfaceReview} 
by gating,\cite{Ueno_gating, Goldman_gating, Goldhaber_gating} 
by forming an interface with a polar perovskite,\cite{Hwang_LAOSTO,OxideInterfaceReview,Caviglia,otherLAOSTO} or 
by placing a $\delta$-doped\cite{deltadoping} layer inside a bulk crystal.
Although 2D electronic systems at 
LaAlO$_3$/SrTiO$_3$ interfaces have received particular attention,\cite{Caviglia,otherLAOSTO} there has
also been important progress with other material systems.\cite{otherOxide}
SrTiO$_3$ two-dimensional electron gases (2DEGs) appear to be strongly correlated when their thermodynamics is probed capacitively\cite{Ashoori_Capacitance} and exhibit both superconductivity\cite{2DEG_super} and 
magnetism,\cite{2DEG_magnetic} sometimes simultaneously.\cite{Ashoori_dual} 
There is at present only a very primitive understanding of the measured properties of these 
potentially interesting 2DEG systems.  The current paper is motivated 
by the view that progress can be accelerated by the development of concrete microscopic models that 
are simplified relative to full {\em ab-initio} electronic structure calculations,\cite{Martin,otherabinitio}
allowing electric properties to be estimated easily and compared with experiment.     



\begin{figure}
\includegraphics[width=8.5cm,angle=0]{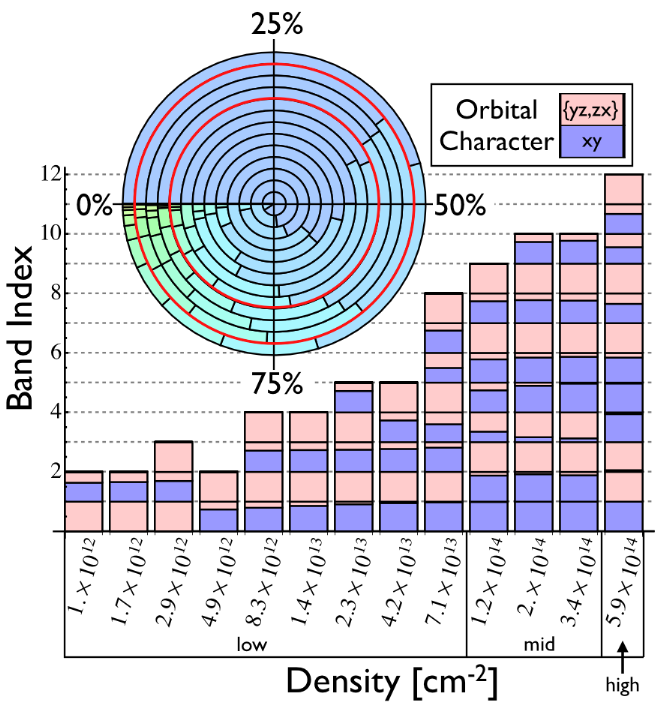} \\
\caption{(Color online) Orbital character at $k_\perp = 0$ of occupied doubly degenerate 2D subbands for a series of 
total areal densities.  The $xy$ and $\{yz,zx\}$ fractions in the orbital character of each band are represented by light blue and light red shading respectively and the band indices are ordered from lowest energy to highest.
Both spin-orbit interactions and tetragonal splitting have been included 
in the band-structure model.  The percentage of the total density associated with a given subband is summarized in the inset piechart, in which the rings from inside to outside correspond to the density values from lowest to highest.
Individual band contributions for a particular density are ordered lowest energy to highest in a clockwise direction.
The red lines in the pie chart separate the low,  mid, and high density regimes identified in the text.  
At the highest total densities, most electrons occupy a small number of strongly 
confined bands.
}
\label{CombinedBandData}
\end{figure}

In this paper we present a model of SrTiO$_3$ 2DEGs that is partly phenomenological
and simplified, but still sufficiently realistic to be predictive.  We focus on
electrostatically gated surface 2DEGs, although our approach applies without 
much change to the case of interface-confined systems.  The same model is 
readily adapted to describe $\delta$-doped 2DEGs inside the STO bulk, 
STO 2DEGs that are modulated by a back gate, and 2DEGs in other $d^{0}$ 
systems, for example KTaO$_3$.  
The model assumes that the itinerant electronic degrees of freedom are derived from the 
SrTiO$_3$ $t_{2g}$ bands.  We use a nearest neighbor tight-binding model to 
describe hopping between TiO$_2$ planes and either tight-binding or $\vec{k} \cdot \vec{p}$ models to 
describe wavefunction variation within TiO$_2$ planes.  
The strength of inter-plane hopping parameters, 
and the values of the heavy and light masses within planes are estimated on the 
basis of recent ARPES\cite{ARPESrefs}
and bulk magnetic oscillation\cite{Stemmer_dHvA} experimental results.
Some aspects of the 2DEG electronic structure are sensitive to the influences of   
spin-orbit coupling and SrTiO$_3$'s low-temperature tetragonal distortion on the host material's 
conduction band, even at the highest 2D carrier densities.  
  
%
%
%
%
%
%
The extremely strong dielectric response of the SrTiO$_3$ lattice
plays a key role in our model at all carrier densities.  Our main results are summarized in Fig.~\ref{CombinedBandData}.
We conclude that unless vertically confined on both sides by vacuum or 
insulating tunnel barriers, SrTiO$_3$ 2DEGs spread  
across a large number of TiO$_2$ planes.  This property is a direct consequence of the host material's 
very large linear dielectric constant, which weakens confinement, and occurs in spite of 
relatively large carrier masses which favor confinement.   At high carrier densities, and hence 
large electric fields, dielectric screening saturates and the 2DEG is mostly confined to 
the first few TiO$_2$ planes.  However a portion of the 2DEG, making a contribution to the 2D 
density that is approximately fixed in absolute terms, still spills over many layers. 
This low-density weakly confined part of the 2DEG can make an important 
contribution to some 2DEG properties.  

Our paper is organized as follows.  In the following section we provide a detailed explanation of the 
model that we use.  We have identified three different density regimes for SrTiO$_3$ surface-bound 2DEGs.
In Secs. III-V we characterize the nature of the 2DEG electronic structure in low ($ n < 
1 \times 10^{14} {\rm cm}^{-2}$), mid ($ 1 \times 10^{14} {\rm cm}^{-2} <  n < 
5 \times 10^{14} {\rm cm}^{-2}$), 
and high ($n > 5 \times10^{14} {\rm cm}^{-2}$) 2DEG carrier density regimes respectively.  Finally in Sec. VI
we summarize our results and speculate on the types of electronic properties that might be 
achievable in SrTiO$_3$ 2DEG systems.  

\section{Model}  
\label{sec:model}

\subsection{$t_{2g}$ tight-binding model} 
SrTiO$_3$ is a non-polar pseudo-cubic band insulator with an electronic gap of $\sim 3.2$ eV separating its 
oxygen $p$-orbital dominated valance band, from its Ti $t_{2g}$-orbital dominated conduction band.
The Ti $e_g$ bands are split by the crystal field and pushed up in energy by $\sim 2$ eV\cite{Perovskites}
relative to the $t_{2g}$ bands and are therefore neglected in our model.   
The conduction band minimum is at the $\Gamma$ point.  The bulk $t_{2g}$ bands are split 
at the $\Gamma$ point, in the first place by spin-orbit interactions which push
one time-reversed pair of $t_{2g}$ bands up by $\sim 18$ meV relative to the other two pairs.  
The two lower energy bands are further split below 110 K by a rotation of the octahedral oxygen environment which is responsible for a tetragonal distortion.\cite{SOTrefs}

The starting point for any phenomenological description of the 
electronic structure of SrTiO$_3$ 2DEGs is an accurate representation of the 
bulk material electronic structure.  Although this material has been
studied for many years, its conduction band minimum was until recently not characterized with an accuracy 
sufficient to model low-carrier density 2DEGs.  
To remedy this, Allen {\em et al.} conducted\cite{Stemmer_dHvA} 
magnetotransport studies on a series of low density electron-doped
MBE-grown samples of SrTiO$_3$ in fields up to 31T
and fit a 6 band $\mathbf{k} \cdot \mathbf{p}$ model of the Ti $t_{2g}$ bands to the 
magnetic oscillation data.  The bulk band parameters used here have been 
taken from that study. 
These experiments suggest that in the absence of spin-orbit coupling the
tetragonal distortion at $4$ K would push the $xy$ bands up by $\sim 3.2$ meV relative to 
$yz$ and $zx$ bands. Together the two-corrections 
fully lift the $t_{2g}$ manifold degeneracy, even in the bulk. (Because of orbital mixing, the spin-orbit (SO) splitting and tetragonal splitting parameters do not directly correspond to the $\Gamma$ point band energies.)

\begin{figure}
\includegraphics[width=8.5cm,angle=0]{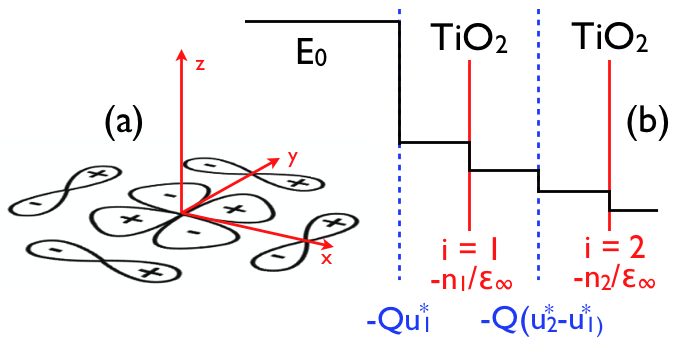} \\
\caption{ (Color online) (a) $p-d$ bonding network for Ti-$xy$ orbital.  The dominant bonding is in-plane $\pi_{pd}$ while the out-of-plane bonding is weak.  (b) Schematic representation of the electric field drop in our model due to lattice screening (dashed blue) and electronic screening (solid red).  Numerical factors have been dropped.}
\label{bondingField}
\end{figure}

Although the $p-d$ oxygen bonding and $e_{g}$ anti-bonding orbitals do not explicitly enter our 
model, they do appear implicitly in the form of the Hamiltonian.  
Consider for example hopping between a Ti $xy$ orbital and its neighboring oxygens within a TiO$_2$ plane (Fig. \ref{bondingField}a).  
Along the x-direction, the dominant bonding is $\pi_{pd}$ through the O-p$_y$ orbital and along the y-direction, $\pi_{pd}$ through the O-p$_x$ orbital.  Overlap with other O-p orbitals is small by symmetry.  
This rule is preserved througout the Ti - O -Ti bonding network.  
For the Ti - $yz$ orbital, $\pi_{pd}$ bonding dominates along the y-direction through O-p$_z$ orbitals.
Bonding along the x-direction vanishes in a Slater-Koster two-center approximation and 
is weak.  Temporarily ignoring the spin-orbit and tetragonal distortion effects, 
these observations suggests a tight binding model for a single isolated layer of the form
\begin{widetext}
\be
\label{HSL}
H_{\sigma}^{SL} = 
\left(
\begin{array}{ccc}
 -2t'cos(k_x) -2t cos(k_y) & 0 & 0 \\
 0 & -2tcos(k_x) -2t' cos(k_y) & 0 \\
 0 & 0 & -2tcos(k_x) -2t cos(k_y) \\
\end{array}
\right)
\left\{
\begin{array}{c}
yz, \sigma \\
zx, \sigma \\
xy, \sigma
\end{array}
\right\} \, ,
\ee 

\noindent
where the cubic lattice constant is used as a length unit, the metal lattice site energies are used 
as the zero-of-energy, $t$ quantifies the 
dominant $\pi_{pd}$ bonding process and $t'$ describes the weaker bonding
process.  
The column on the right specifies the orbital representation used for this 
Hamiltonian matrix.
Hopping terms that couple different $t_{2g}$ orbitals are allowed\cite{Bistritzer} from a 
symmetry point of view.  However, Allen {\em et al.} were unable to distinguish this mixing parameter from zero in their recent analysis of SdH data.\cite{Stemmer_dHvA}  We therefore ignore these processes in our model.   
For lower carrier densities it is sometimes convenient to use a simplified version of this 
model in which we expand Eq. \ref{HSL} around the 2D $\Gamma$ point.  We find that for 
2D wavevectors that are small compared to Brillouin-zone dimensions 
\be
\label{HSLcont}
H_{\sigma}^{SL} = 
\left(
\begin{array}{ccc}
 \epsilon_{yz,0} + t' k_x^2 + t k_y^2 & 0 & 0 \\
 0 & \epsilon_{zx,0} + t k_x^2 + t' k_y^2 & 0 \\
 0 & 0 & \epsilon_{xy,0} + t k_x^2 + t k_y^2 \\
\end{array}
\right)
\left\{
\begin{array}{c}
yz, \sigma \\
zx, \sigma \\
xy, \sigma
\end{array}
\right\} \, ,
\ee 

\end{widetext} 

\noindent
where $\epsilon_{yz,0} = \epsilon_{zx,0} = -2 t - 2 t' $ and $\epsilon_{xy,0} = -4t$. 
We use this low density form for the planar Hamiltonian for most of the calculations presented below. The 
more general tight-binding model must be used when 2D carrier densities are large and 
confinement is strong, and can be adopted when required without essential 
complication.

In the same representation, adjacent 2D layers are coupled by an 
interlayer hopping term of the form 

\be
\label{Hij}
H_{\sigma}^{C} = 
\left(
\begin{array}{ccc}
t & 0 & 0 \\
 0 & t & 0 \\
 0 & 0 & t' \\
\end{array}
\right)
\left\{
\begin{array}{c}
yz, \sigma \\
zx, \sigma \\
xy, \sigma
\end{array}
\right\} \, .
\ee 

\noindent
Here the symmetry of the bonding network has again been employed to note that 
the $xy$ orbital has the weaker interlayer coupling, $t'$.  
Because $t'$ is expected to be substantially smaller than $t$, the  
$xy$ bands in single-layer 2DEGs are pulled down by $\sim 2 t$ at the $\Gamma$ point 
relative to the $yz$ and $zx$ bands.
In the bulk limit, on the other hand, the three bands are degenerate because each has two strong 
hopping and one weak hopping direction.  
Any amount of confinement in the $\hat{z}$ direction pushes the bottom
of the $\{yz,zx\}$ bands up relative to the $xy$ band and leads to orbital polarization.

For low-carrier densities on-site (${\bm k}$-independent) terms due to 
tetragonal distortions and spin-orbit coupling must be included.\cite{Bistritzer}
The tetragaonal distortion is represented by a parameter 
$\Delta_T$ which characterizes the difference in site energy between $xy$ and $\{yz,zx\}$-orbitals,
and spin-orbit coupling by an interaction strength parameter $\Delta_{SO}$. 
The distortion Hamiltonian is  

\be
\label{Tii}
T_{\sigma} = 
\left(
\begin{array}{ccc}
0 & 0 & 0 \\
 0 & 0 & 0 \\
 0 & 0 & \Delta_T \\
\end{array}
\right)
\left\{
\begin{array}{c}
yz, \sigma \\
zx, \sigma \\
xy, \sigma
\end{array}
\right\} \, ,
\ee 
and the spin-orbit Hamiltonian, modeled in an atomic approximation, is 

\be
\label{SO}
H^{SO}  = \frac{\Delta_{SO}}{3}
\left(
\begin{array}{cccccc}
0 & i & 0 & 0 & 0 & -1 \\
-i & 0 & 0 & 0 & 0 & i \\
0 & 0 & 0 & 1 & -i & 0 \\
0 & 0 & 1 & 0 & -i & 0 \\
0 & 0 & i & i & 0 & 0 \\
-1 & -i & 0 & 0 & 0 & 0 \\
\end{array}
\right)
\left\{
\begin{array}{c}
yz, \uparrow \\
zx, \uparrow \\
xy, \uparrow \\
yz, \downarrow \\
zx, \downarrow \\
xy, \downarrow \\
\end{array}
\right\}.
\ee 

Our model for the electronic structure of SrTiO$_3$ 2DEGs combines the single-particle model explained above
with a Hartree approximation for electron-electron interactions.  The external 
electric field which produces surface confinement is screened by carriers and by lattice 
relaxations of the partially ionic SrTiO$_3$ crystal.  In SrTiO$_3$ lattice screening is strong and 
non-linear and plays a subtle and essential role in confinement.
 
\subsection{Lattice relaxation model}

The exceptionally strong and temperature-dependent linear dielectric response of bulk SrTiO$_3$ is
associated with a soft optical phonon mode in which 
positively charged Sr and Ti atoms move in opposition to the negatively charged oxygen octahedra.
Displacement of this mode in response to an external electric field produces screening. 
Because the mode is extremely 
soft only near the center of the Brillouin-zone,\cite{Cowley} it responds strongly 
only when a large external field persists over several TiO$_2$ layers.
In addition this screening response is very non-linear, saturating at very large electric fields.
Since the reduction in electric field is proportional to the phonon mode displacement,
saturation occurs because the phonon mode is anomalously soft only for small displacements.\cite{Cowley}
In an attempt to capture this behavior qualitatively, we use a simple model of 
lattice relaxation which focuses on the soft-mode only.  We write the lattice energy as 
%
%

\be
\label{EnergyLattice}
U = \frac{1}{2}\sum_{i,j} u_i K_{i,j} u_j - Q \sum_{i} E_i u_i + \frac{\gamma}{4} \sum_{i} u_i^4 
\ee

\noindent
where $u_i$ is the displacement of the soft-mode on the $i$-th lattice site,
$E_i$ is the average electric field in the $i^{th}$ cell, $Q$ is an effective charge defined in terms 
of the polarization-density per unit soft-mode displacement, and  
$\gamma$ is a parameters chosen to capture the non-linearity 
of the dielectric response as discussed further below.  Here $K_{i,j} = K_{i-j}$ is the dynamical matrix at 
2D wavevector ${\bm q_{\perp}}=0$.  We fit $K_{ij}$ to the soft-mode 
phonon dispersion using a form with a local on-site contribution and a 
Gaussian non-local contribution.  In momentum space, this takes the form,
\be
\label{phononFit}
K(q,G) = (2\pi)^2 \mu \left[ f_0^2 - f_1^2 e^{-\frac{\alpha_1^2}{2} (q+G)^2}  - f_2^2 e^{-\frac{\alpha_2^2}{2} (q+G)^2} \right]
\ee

\noindent
where $q$ is the lattice momentum, 
$G$ is a reciprocal lattice vector, and 
$\mu = 24amu$ is the appropriate
reduced mass for the Ti atom moving opposite the oxygen octahedra.
The parameters $f_0$, the strength of the on-site term, and $f_1$ and $\alpha_1$ were 
chosen to reproduce the measured phonon dispersion.  Because no low temperature phonon 
data exists in the literature, $f_2$ and $\alpha_2$ have been added to capture the low temperature dielectric response of the bulk material.  By minimizing Eq. \ref{EnergyLattice} in the absence of an electric field and evaluating Eq. \ref{phononFit} at $q=0$ we find that 

\be
\gamma = \left[ \frac{2 \pi (f_0-f_1-f_2)}{u_{NL}} \right]^2
\ee

\noindent
where $u_{NL}$ is the mode displacement at which non-linear dielectric response is seen.
(See the discussion below).

Because the relative displacements of all atoms are known, only a single displacement coordinate, $u_i$, 
is needed to describe the response of the unit cell to perturbations along the principle crystal axes. 
Given the electric field in each cell of the crystal, Eq. \ref{EnergyLattice} can be minimized to find the 
appropriate set of displacements, $\{u_i^*\}$.  We define the three-dimensional 
polarization density of the SrTiO$_3$ as 

\be
\label{polarization}
P_i = \frac{1}{a^3} Q u_i^* \left[ \Theta(z_i -a/2) - \Theta(z_i + a/2) \right]
\ee

\noindent
where $z_i$ is the location of the TiO$_2$ layer of interest, $a$ is the lattice constant of the crystal, and $\Theta(z)$ in the Heaviside function.  The precise way in which the polarization density is mapped onto our lattice model is immaterial on length scales larger than a lattice constant. 
To find the effective charge parameter we use the standard definition of the screened electric field and linear dielectric constant,

\be
\label{ScreenedField}
E = E_0 + 4\pi P \approx \epsilon E_0.
\ee

\noindent
After minimizing Eq. \ref{EnergyLattice} in the linear, bulk limit and using the definition of the polarization from Eq. \ref{polarization}, we find that,

\be
\label{EffectiveCharge}
Q =\sqrt{   \frac{\mu \omega_1^2 }{4 \pi}(\epsilon-1) }.
\ee

\noindent
To make contact with the measured properties of the bulk material in a straightforward way, we use $90K$ values for the phonon dispersion \cite{Cowley} and dielectric constant. \cite{Neville_dielectric}  In terms of model parameters $\omega_1 = 2\pi (f_0 - f_1)$ and the $90K$ dielectric constant are given in Table 1.  With this we find $Q=8.33e$, a value comparable to those used in models of this type for bulk SrTiO$_3$. \cite{Cowley}

The electric field in Eq.~\ref{EnergyLattice} can be found by solving

\be
\label{Efield}
\nabla \cdot E(z) = -\frac{4 \pi e}{\epsilon_\infty} \sum_i n_i \delta(z-z_i) + 4 \pi \sum_i \nabla \cdot P_i,
\ee
with the boundary conditions that $ E(-\infty) = E_0$ and $E_{\infty}=0$.
The electric field boundary conditions are discussed below.   In Eq.~\ref{Efield}, $e$ is the electron charge, $\epsilon_\infty$ is the high frequency dielectric constant 
due to electronic screening, and $n_i$ is the number density of itinerant electrons in TiO$_2$ layer $i$.
Both lattice relaxation and conduction band charge accumulation screen the external electric field.  This is represented pictorially in Fig. \ref{bondingField}b.

\subsection{Electric Field Boundary Conditions}  

In the calculations presented below we assume that the electric field above the surface of the SrTiO$_3$ has been set 
experimentally either by gating or by forming an interface with a polar dielectric.\cite{polarizationcatastrophe}
In the latter case $E_0$ is ideally set by the polarity of the material, but can also be influenced by 
surface reconstructions or other detailed material considerations that can be sensitive to uncontrolled 
aspects of growth.  Because we have gated systems in mind, we consider that $E_0$ can be varied experimentally 
over a broad range.  In this calculation we set the electric field below the SrTiO$_3$ 2DEG, $E_{bulk}$, to zero, assuming that the sample lies on a grounded metallic substrate.  (If the SrTiO$_3$ sample was thin, a conducting layer under the sample could be used as a gate and $E_{bulk}$ could be varied.)  
By integrating the Poisson equation (Eq.~\ref{Efield}) and noting that
the lattice relaxation contribution to $E(z)$ vanishes far below the surface when $E_{bulk} \to 0$, we conclude that 
the 2DEG density is fixed by $E_0$ alone: $n_{T} = \sum_i n_i = \epsilon_\infty E_0/4\pi e$.  
We can therefore replace this parameter by the total 
2DEG density $n_{T}$ and present results as a function of that parameter. 

We incorporate the layer-dependent electric potential contribution to the Hamiltonian by integrating $E(z)$ across the 
2DEG to obtain a layer-dependent potential $V_i$ which must be 
determined self-consistently along with the 2DEG density-distribution and 
the soft-mode displacement field.  With this, the Hamiltonian of the system becomes,

\be
\label{Ham}
H = \sum_{<i,j>} \vec{c}_i{\,^\dagger}  H^{C} \vec{c}_j + \sum_i  \vec{c}_i^{\,\dagger} \left( H^{SL} + T + H^{SO} + V_i\right) \vec{c}_i
\ee

\noindent
where the double sum in the first term is over neighboring layers.  In Eq. \ref{Ham}, we work in the representation $\vec{c} = \{ {c_{xy,\uparrow},c_{yz,\uparrow},c_{zx,\uparrow},c_{xy,\downarrow},c_{yz,\downarrow},c_{zx,\downarrow}} \}$ so that
$H^{SO}$ has the form of Eq. \ref{SO}.


The layer resolved density $n_i = <\vec{c}_i^\dagger \vec{c_i}>$ is calculated from Eqs. (\ref{Efield},\ref{Ham}) and minimization of Eq. \ref{EnergyLattice}.  We have carried these self-consistent field calculations to convergence 
over a wide range of densities for a system that is $60$ unit cells wide.
Because of the long tail in the density-distribution discussed at length below, it is difficult to achieve self-consistency and 
we were forced to mix in no more than $\sim 1\%$ of new results in the iterative update of the charge density.
Although the model described in this section is crude in some respects, certainly crude compared to 
{\em ab initio} electronic structure calculations with full lattice relaxation, it is strongly motivated by the 
cumbersome character of the fully microscopic calculations under these circumstances.  The model could be made
more quantitative by being bench-marked against {\em ab initio} calculations or, perhaps more reliably, by 
comparison with experiment.  

\begin{table}
\begin{tabular}{| l | l | l |}

\hline
\multicolumn{3}{| c |}
{Model Parameters} \\
\hline

Lattice Constant & $a$ & $3.904$ \AA \\ \hline

\multirow{4}{*}
{Electronic Parameters} & $t$ & $236meV$ \\
 & $t'$ & $35meV$ \\
 & $\Delta_{SO}$ & $18meV$ \\
  & $\Delta_{T}$ & $3.2meV$ \\ 
  \hline
  
\multirow{5}{*}
{Dielectric Response} & $ \epsilon_0$ & $24408$ \\
 & $\epsilon_1$ & $1340$ \\
 & $\epsilon_\infty$ & $5.5$ \\
 & $Q$ & $8.33 e$ \\
 & $u_{NL}$ &  $0.0034$\AA \\ 
 \hline
 
\multirow{5}{*}
{Dynamical Matrix} & $f_0$ & $4 \times 10^{12} c/s$ \\
& $f_1$ & $2.73 \times 10^{12} c/s$ \\
& $f_2$ & $0.97 \times 10^{12} c/s$ \\
& $\alpha_1$ & $1.15 a$ \\
& $\alpha_2$ & $5 a $ \\ 
\hline

\end{tabular}
\label{Table}
\caption{Parameters used in the current study.  The electronic structure parameters have been taken from Ref. \onlinecite{Stemmer_dHvA},  
while $\epsilon_\infty$, $\epsilon_1$ and $\epsilon_0$ were taken from Refs. \onlinecite{Cowley} $\&$ \onlinecite{Neville_dielectric}.}
\end{table}

%
%

\section{Low Carrier Densities $1 \times10^{14}{\rm cm}^{-2} < n_{T}$} 

For the circumstance considered here
the total carrier density is proportional to the electric field just above the SrTiO$_3$ surface and 
the largest internal electric fields are closest to the surface.  
We define the low-density regime by the requirement that the largest electric fields are 
smaller than the scale at which non-linear screening becomes important.  This field scale is set by the 
model parameters $u_{NL}$ and $\epsilon_{\infty}$, which can be determined approximately from experiment.  
We have estimated $u_{NL}$ by comparing Eq. \ref{polarization} with the deviation from linear response seen in
the polarization of bulk STO crystals.\cite{Neville_dielectric}  
This value is listed along with other model parameters in Table I.  
The model parameters we have chosen reflect the estimate that non-linear screening becomes important for carrier densities larger than $\sim 1 \times 10^{14} cm^{-2}$.

\begin{figure}
\includegraphics[width=8.5cm,angle=0]{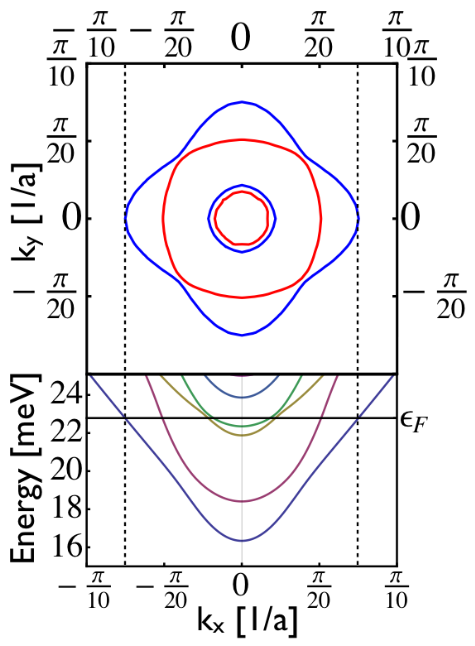} \\
\caption{(Color online) (a) Fermi surface and 2D band structure for $n_T = 8.3 \times 10^{12}$. The dominant orbital character at the 2D $\Gamma$ point is represented for each band by the color of the 
bands and Fermi lines - blue and red for $xy$ and $\{yz,zx\}$, respectively.  
The zero of energy is set to the potential minimum in the first layer.  
The Fermi energy is indicated by a solid (black) horizontal line.
} \label{LowDensityBands}
\end{figure}

For linear screening some 2DEG properties are similar to those of covalent semiconductor 2DEGs 
and can be estimated following the same lines as in Stern's pioneering study of Si/SiO$_2$ MOSFET 
2DEGs.\cite{Stern}  In particular the confinement length scale $w$ can be crudely 
estimated by equating the quantum confinement kinetic energy 
cost and the confinement electric potential scale.  Neglecting numerical factors we therefore set  
\be
\label{lengthscale} 
\frac{\hbar^2}{mw^2} \sim \frac{e E_{0} w}{\epsilon} \sim \frac{4\pi e^2 n_{T} w }{\epsilon} 
\ee 
to obtain 
\be
w \sim \Big(\frac{\hbar^2 \epsilon }{ 4 \pi m e^2 n_{T}} \Big)^{1/3}. 
\ee  
In the linear screening regime the confinement length scale decreases quite slowly with the total 2DEG density.  
The hopping parameters of Table I can be converted to effective masses for the $t_{2g}$ bands;
the light mass that describes the vertical confinement of the most poorly confined $\{yz,zx\}$ bands is $\sim m_0$ where 
$m_0$ is the bare electron mass.  When combined with the extremely large low-temperature 
bulk dielectric constant of SrTiO$_3$ ($\epsilon \sim 25000$), we estimate that $w$ is close $\sim 50$  
SrTiO$_3$ unit cells even at the top end of the low-density regime.  We therefore expect that the hard wall at 
60 unit cells used in our calculations influences our numerical results.  The main point of these 
qualitative considerations is that we should expect weak surface confinement
at low carrier densities because of very strong dielectric screening. 

In Figs.~\ref{LowDensityBands} we illustrate a 
typical 2D band structure in the low density regime.  
Here the bottom band is beginning to reflect the increase in  
$xy$ character expected from confinement, 
and the small size of the subband splittings is in qualitative agreement with
the estimated scale of size-quantization energies: 
\be
\frac{\hbar^2}{mw^2} \sim 10^{-4} \, {\rm eV}.
\ee   
The small subband splittings imply that the 2DEG is 3D in character 
unless temperatures are low and disorder extremely weak.  The vertical spread of the 
2DEG is expected to get smaller, and the subband splitting larger with increasing 
temperature as the dielectric constant value decreases.\cite{Neville_dielectric}

\begin{figure}
\includegraphics[width=8.5cm,angle=0]{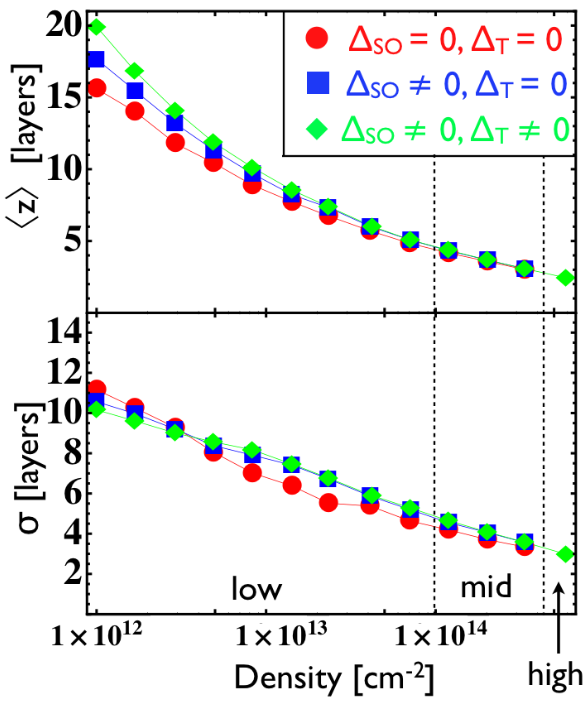} \\
\caption{(Color online) Average separation from surface, $\langle z \rangle$ and standard deviation $\sigma$
of the electron distribution across layers as a function of total density, and its dependence 
on SO splitting and the tetragonal distortion.  The tendency of SO coupling and tetragonal splitting to 
weaken surface confinement is suppressed when densities reach the mid range.  
When confinement energy scales are not strong enough to overcome the tetragonal distortion, 
SO and tetragonal-splitting induced hybridization decrease the spread of the $\{yz,zx\}$ bands (see text).}
\label{Cumulants}
\end{figure}

\begin{figure}
\includegraphics[width=8.5cm,angle=0]{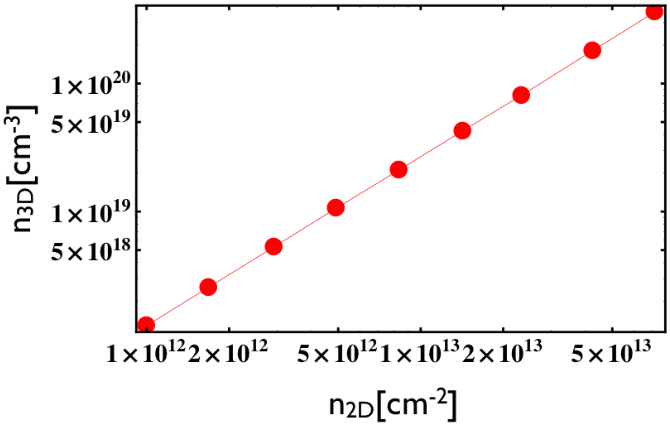} \\
\caption{(Color online) Calculated 3D density in the low density regime.  The relationship between
2D density and average 3D density follows a $4/3$ power law 
that is consistent with Eq. \ref{lengthscale} is correct.}
\label{n3d}
\end{figure}

Low carrier density properties are strongly influenced by spin-orbit coupling
which hybridizes the $t_{2g}$ basis states and induces a splitting at the $\Gamma$ point
in the bulk.  One effect of spin-orbit coupling is to weaken the 2DEG surface confinement by hybridizing 
$xy$ bands with $yz$ and $zx$ bands that have smaller masses perpendicular to the surface.  
Spin-orbit induced hybridization allows the $xy$ subbands to communicate between layers through 
their $\{yz,zx\}$ admixtures which are less easily confined.  
In the low density regime, spin-orbit splitting is pronounced enough to change the dominant orbital character of the 2D subbands.  

The tetragonal distortion increases the site energy of the $xy$ band - further enhancing the 
role of the less confined $\{yz,zx\}$ components.  (We have assumed that 
the tetragonal axis is parallel to the surface normal, as expected near a surface.)  
Initially, confinement energy scales are weak compared to the tetragonal splitting energy.
As the carrier density and the energetic width of the occupied states increase, spin-orbit
coupling becomes less important and the $xy$ fraction of the lowest energy most highly occupied band 
increases.  (See Fig.~\ref{CombinedBandData}.) The influence of the spin-orbit and tetragonal splittings on the 
spatial distribution of electrons is summarized in Fig.~\ref{Cumulants}.  Estimating 3D densities using $n_{3D} = n_{2D}/ \langle z \rangle$ where 
$\langle z \rangle$ is taken from Fig. \ref{Cumulants} and $n_{2D}$ is the total density in the 
linear screening spatial region, we find a power law of $4/3$.  
This suggests that the qualitative estimate of Eq. \ref{lengthscale} is accurate when screening is linear.

\begin{figure}
\includegraphics[width=8.5cm,angle=0]{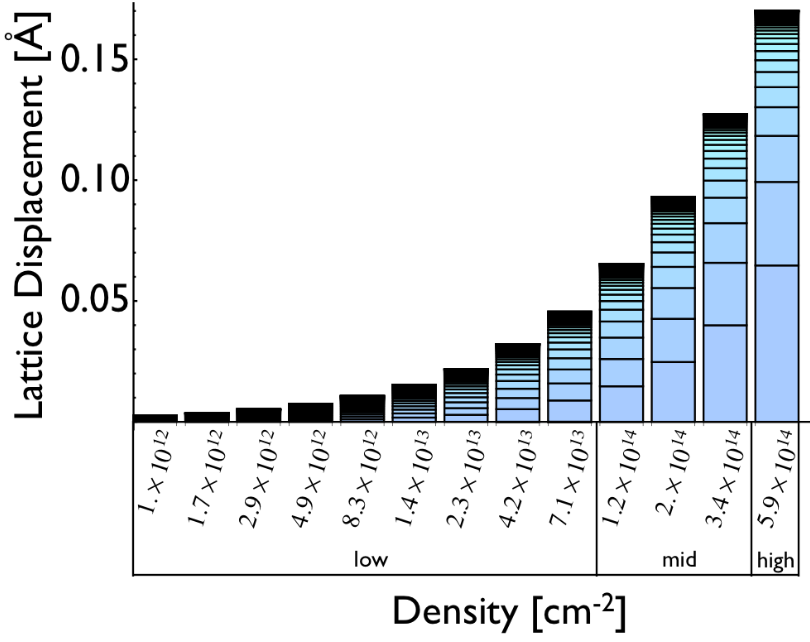} \\
\caption{ (Color online) Lattice displacement as a function of total 2D density.  For each density, the lattice displacement of 
layer $i$ (counting from the surface) is plotted as the height of the $i^{th}$ bar segment (counting from the bottom).
For low-densities, the lattice displacements are small and in the linear regime.  For mid-range densities, lattice displacements are suppressed by non-linear screening effects near the surface.  Weaker lattice screening results in stronger 
confinement, larger 2D subband separations, and fewer occupied 2D subbands.   
}
\label{Lattice}
\end{figure}

\section{Mid-range densities:  $1 \times 10^{14} {\rm cm}^{-2} < n < 5\times10^{14} {\rm cm}^{2}$} 

We define the mid-range of densities as that in which lattice screening is markedly reduced 
because of non-linear dielectric screening (see Fig. \ref{Lattice}.)  Because the electric field is larger
closer to the surface, non-linear screening is more important there.  The strong surface 
electric fields cause a large fraction of the total electron density be confined close to the surface, 
and size-quantization effects to increase much more rapidly with carrier density than 
would be suggested by 
Eq.~\ref{lengthscale}.  
Even though a substantial fraction of the total charge density starts to become confined within the 
top few layers, there is still a wide tail in the density distribution in the spatial region over
which the external electric field has been reduced to a value less than 
$\sim 10^{14} {\rm cm}^{-2}\epsilon_\infty/(4 \pi e) $ so that the screening is 
locally linear.  In our model the non-locality of these screening properties is 
set by the width in momentum space of the long-wavevector limit of the soft mode.  
In our numerical calculations, this low-density quasi 3D regime is influenced by our 
hard-wall cut-off of the 2DEG beyond a width of 60 unit cells.   

\begin{figure}
\includegraphics[width=8.5cm,angle=0]{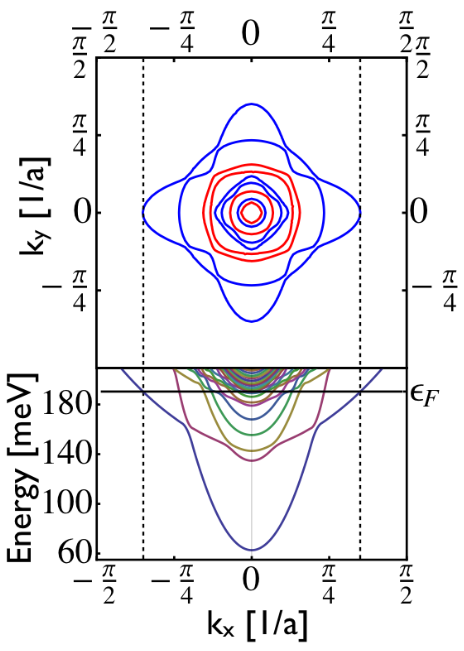} \\
\caption{(Color online) (a) Fermi surface and 2D band structure for $n_T = 2.0 \times 10^{14}$. The dominant orbital character of a band at the 2D $\Gamma$ point is represented by line color with blue and red 
indicating $xy$ and $\{yz,zx\}$, respectively.  The zero of energy is set to the potential minimum in the first layer.  
The Fermi energy is represented by a solid (black) horizontal line.
Although the separation between the lowest energy 2D subbands is large, many low density subbands with 
small energy separations are
still present near the Fermi energy.
} 
\label{MiddleDensityBands}
\end{figure}

As was the case in the low density regime, the inclusion of SO coupling and the tetragonal distortion
alters the the orbital character of the lowest energy band and 
decreases its surface confinement.  Their influence is reduced compared to the low-density regime however.
As illustrated in Fig.~\ref{MiddleDensityBands} 
we find that, at the 2D $\Gamma$ point, the two lowest bands are dominantly {\em xy} in character and that the 
next occupied subbands are $\{yz,zx\}$ in character.  Although the number of 2D subbands has increased significantly, only a few are needed to account for the most strongly confined part of the density (see the inset of Fig. \ref{CombinedBandData}).

\begin{figure}
\includegraphics[width=8.5cm,angle=0]{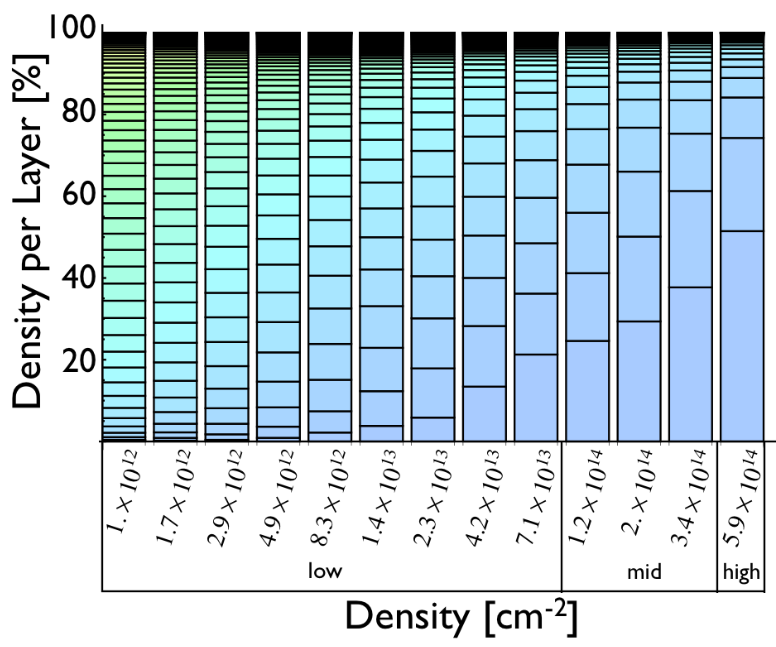} \\
\caption{(Color online) Layer resolved density (as a percent of total density) 
as a function of the total density.  For each density each segment represents the percent of the total density in the corresponding TiO$_2$ layer, starting with the first layer and moving upward.  For low densities the 2DEG is spread over many layers.  Above $n_T \sim 10^{14} cm^{-2}$, the confinement becomes pronounced.  In the high density regime, more than $50\%$ of the total density is confined within the first layer due mainly to reduced lattice screening at 
large electric fields.
} \label{density}
\end{figure}

\section{High carrier densities - above $5\times10^{14}$}

\begin{figure}
\includegraphics[width=8.5cm,angle=0]{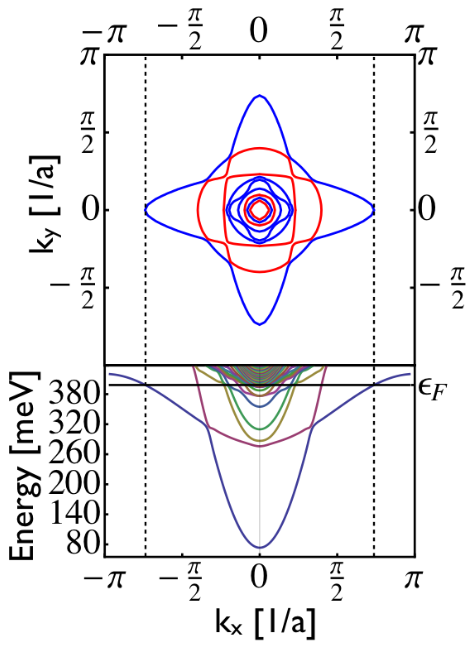} 
\caption{(Color online) (a) Fermi surface and 2D band structure for $n_T = 5.9 \times 10^{14}$. The dominant orbital character of the band at the 2D $\Gamma$ point is represented by line color with blue and red for $xy$ and $\{yz,zx\}$, respectively.  The zero of energy is set to the potential minimum in the first layer.  The Fermi energy is represented by a solid (black) horizontal line.  The 2D subband splitting is $\sim 200meV$ - becoming comparable to the single layer model of Eq. \ref{HSL}.  Near the Fermi energy, many subbands are present with $\sim meV$ splitting, which contribute to the carrier density 
far from the surface.  
} 
\label{HighDensityBands}
\end{figure}

The 2DEG electronic structure simplifies again in the high-density limit, which we define as the limit with 
more than half of the total density in the first TiO$_2$ layer - see Fig.~\ref{density}.  For large electric fields, and therefore 
large carrier densities, lattice screening is irrelevant near the surface.  Because of the relatively large conduction
band masses, compared to typical covalent semiconductor cases, and the much stronger electric fields 
at these carrier densities, surface confinement occurs on an atomic length scale.  The $\Gamma$-point splitting becomes comparable to the single layer limit of Eq. \ref{HSLcont} - see Fig. ~\ref{HighDensityBands}.  While the SO coupling leads to hybridization and a decrease in confinement that mainly affects the quasi 3D tail of the electronic distribution, the tetragonal distortion does not have a noticeable effect.  In this regime,  if we neglect the quasi 3D tail region, there are only a few spin-degenerate 2D subbands contributing to the density.  At the 2D $\Gamma$ point, the first three bands are dominantly $xy$ and $\{yz,zx\}$ - going from low to high energy.  For the high-density regime, the tight-binding model of Eq. \ref{HSL} must be used.

\section{Summary and Discussion}   

Using a simplified tight-binding model for the $t_{2g}$ bands,
we find that non-linear screening plays an essential role in determining the electronic properties of 
surface confined 2DEGs in grounded SrTiO$_3$.  
For low-density ($n_T < 10^{14} {\rm cm}^{-2}$) 2DEGs, electrons are distributed over many layers because
surface confinement is weakened by the 
host material's extremely large linear dielectric constant.  
In the mid and high density regimes, a low density tail is still present over 
50 or more layers but a high density region 
emerges over the first few atomic layers.  
Although many 2D subbands are always present in the mid and high density regimes, the few 
lowest energy subbands which contain strongly confined orbitals account 
for most of the total density (see Fig. \ref{CombinedBandData}).
Subbands that have substantial $\{yz,zx\}$ orbital character are present at all densities in the grounded
configuration investigated in this study.  The presence of many subbands 
with different orbital character at the Fermi level
suggests that the interpretation of transport properties in these systems is not likely to be straightforward.  

SrTiO$_3$ is well known for exhibiting superconductivity in the bulk \cite{Schooley}
where it appears over a broad range of carrier densities from $\sim 10^{18} {\rm cm}^{-2}$ to more than $\sim 10^{20}{\rm cm}^{-2}$ and has a maximum value $\sim 400$mK.\cite{Schooley}   In one study of gated SrTiO$_3$ 2DEG systems, \cite{Ueno_gating} with Hall densities in the range from $\sim 10^{13} {\rm cm}^{-2}$ to $\sim 10^{14}  {\rm cm}^{-2}$, the superconducting transition temperature initially increased with carrier density but decreased at 
the higher densities.  Referencing to Fig. \ref{n3d} for 3D densities associated with the weakly confined tail, we find that the
measured surface 2DEG $T_c$'s compare well with values reported for bulk systems. \cite{Schooley}  
In another study, \cite{Goldman_gating} superconductivity was seen only at a Hall density of $3.9\times 10^{14}{\rm cm}^{-2}$ with a transition temperature of near $\sim 400mK$.  Because of experimental limitations superconductivity was not seen at other densities but could have been present at lower transition temperatures. \cite{Goldman_gating} 
In LaAlO$_3/$SrTiO$_3$ systems the reported 2DEG densities are in the 
low density range.  It is therefore not surprising that the measured $T_c$ 
values are correspondingly suppressed.\cite{2DEG_super, Ashoori_dual}  
(The carriers found in LaAlO$_3/$SrTiO$_3$ 2DEG systems are thought to be induced by 
a polarization discontinuity, although the small value of the measured 2D 
carrier densities is not completely understood.) 
We conclude that existing studies are consistent with surface 2DEG and bulk superconductivity in SrTiO$_3$ 
having a common origin.  

The strongly confined portion of the electron distribution in the mid and high density regimes has significant $\{yz,zx\}$ character. (See Figs. \ref{MiddleDensityBands} $\&$ \ref{HighDensityBands}). 
The increase in the density of states associated with these heavy 2D bands could account for the observed ferromagnetism,\cite{2DEG_magnetic,Ashoori_dual} if it is describable by a Stoner criterion.
The spatial separation between the strongly confined and the low density tail portions of the 2DEG
distribution may account for the coexistence of superconductivity and magnetism seen in some systems.\cite{Ashoori_dual}
This scenario should be compared with one in which superconductivity and magnetism both occur in 
strongly confined subbands; the two pictures should be experimentally distinguishable because of the strong 
influence of magnetism and spin-orbit coupling on superconducting properties\cite{Lee} in the spatially
coincident case. 

The low density tail is a consequence of the property that the electric field vanishes far 
from the surface of a grounded system with a surface-bound 2DEG.  The non-linear screening 
properties that we have discussed imply that a back-gate applied to the 
surface 2DEG to increase the strength of the electric field strength deep below the surface
will have an exaggerated impact on the low-density tail of the distribution function and on the corresponding  
closely-spaced 2D subbands near the Fermi surface.  A gate electric field with 
strength $\sim 10^{14}cm^{-2} \epsilon_\infty/4\pi e$ should essentially eliminate the tail region.  
Our prediction that superconductivity is associated with the low density tail can 
therefore be tested by back gating which  
should suppress and eventually eliminate superconductivity\cite{Caviglia} without 
having a large influence on magnetism.  Irrespective of the reliability of these predictions,
it seems clear that studies of the electronic properties in dual-gated samples could be 
quite informative in building up a confident understanding of 2DEG properties.  

\acknowledgements

This work was supported by the National Science Foundation under grants DGE-0549417 and DMR-1122603
and by the Welch foundation under grant TBF1473.  
The authors acknowledge valuable conversations with Jim Allen, Ray Ashoori, Harold Hwang, and Susanne Stemmer.

Ê

\end{document}